# Above-room-temperature intrinsic ferromagnetism in ultrathin van der Waals crystal $Fe_{3+x}GaTe_2$

Gaojie Zhang[1,2,†], Jie Yu[1,2,†], Hao Wu[1,2], Li Yang[1,2], Wen Jin[1,2], Bichen Xiao[1,2], Wenfeng Zhang[1,2,3], Haixin Chang[1,2,3,*]

[1]State Key Laboratory of Material Processing and Die & Mold Technology, School of Materials Science and Engineering, Huazhong University of Science and Technology, Wuhan 430074, China.

[2]Wuhan National High Magnetic Field Center and Institute for Quantum Science and Engineering, Huazhong University of Science and Technology, Wuhan 430074, China.

[3]Shenzhen R&D Center of Huazhong University of Science and Technology, Shenzhen 518000, China.

[†]Theses authors contributed equally to this work.

[*]Corresponding authors. E-mail: hxchang@hust.edu.cn

**ABSTRACT:** Two-dimensional (2D) van der Waals (vdW) magnets are crucial for ultra-compact spintronics. However, so far, no vdW crystal has exhibited tunable above-room-temperature intrinsic ferromagnetism in the 2D ultrathin regime. Here, we report the tunable above-room-temperature intrinsic ferromagnetism in ultrathin vdW crystal $Fe_{3+x}GaTe_2$ (x = 0 and 0.3). By increasing the Fe content, the Curie temperature ($T_C$) and room-temperature saturation magnetization of bulk $Fe_{3+x}GaTe_2$ crystals are enhanced from 354 to 376 K and 43.9 to 50.4 emu·g$^{-1}$, respectively. Remarkably, the robust anomalous Hall effect in 3-nm $Fe_{3.3}GaTe_2$ indicate a record-high $T_C$ of 340 K and a large room-temperature perpendicular magnetic anisotropy energy of $6.6 \times 10^5$ J·m$^{-3}$, superior to other ultrathin vdW ferromagnets. First-principles calculations reveal the asymmetric density of states and an additional large spin exchange interaction in ultrathin $Fe_{3+x}GaTe_2$ responsible for robust intrinsic ferromagnetism and higher $T_C$. This work opens a window for above-room-temperature ultrathin 2D magnets in vdW-integrated spintronics.

**Main text**

In the past several years, two-dimensional (2D) magnetic crystals have shown great potential in low-dimensional physics and spintronics[1-7]. Their van der Waals (vdW)-layered structures and intrinsic ferromagnetic/antiferromagnetic ground states at finite temperatures provide a promising platform for exploring quantum phenomena in the 2D limit[8,9], and further offer possibilities for atomically-thin magneto-electrical and magneto-optical devices with atomically-smooth interfaces[5-7]. In general, magnetic transition temperature will be strongly suppressed by thermal fluctuations when the thickness of vdW magnets down to 2D regime[2,9-12], while magnetic anisotropy can counteract thermal fluctuations for stabilizing 2D magnetic order[1]. Thus, bulk vdW ferromagnetic crystals should have sufficiently high Curie temperature ($T_C$) and large magnetic anisotropy in order to maintain robust ferromagnetism in few layers. However, all known vdW ferromagnets present low $T_C$ or poor room-temperature ferromagnetic properties[1,2,9,13], which greatly impeded their practical applications. Previous efforts have tuned the $T_C$ and magnetic anisotropy of vdW ferromagnets by external ionic liquid gating, but the introduction of liquid has limited the application scenarios[14-16]. Therefore, it is still elusive and very challenging to simultaneously achieve tunable above-room-temperature intrinsic ferromagnetism and large magnetic anisotropy in an ultrathin 2D vdW crystal.

Recently, the realization of above-room-temperature $T_C$ (~367 K for bulk crystal and ~350 K for 9.5-nm nanosheet) and large perpendicular magnetic anisotropy (PMA) in vdW intrinsic ferromagnetic crystal $Fe_3GaTe_2$ has been a milestone[3], leading to the all-vdW-integrated spintronic devices at room temperature[17-22]. $Fe_3GaTe_2$ shows highly spin-polarized Fermi surfaces[23] and large magnetic interactions with electron correlations[24,25]. However, previous studies have demonstrated that the $T_C$ of $Fe_3GaTe_2$ begins to rapidly decrease when its thickness down to few layers[11,12]. Moreover, considering the thermal effects of actual electronic devices during operation, further improving the intrinsic ferromagnetism in $Fe_3GaTe_2$ is crucial for enhancing its application potential in 2D spintronics. Compared with magnetic modulation strategies such as electrical methods[14,26,27] and heterostructure engineering[28], atomic doping is an effective method that does not introduce Joule heating effect and additional interfaces[29]. However, magnetic heterodopants (such as Co or Ni) will alter the magnetic state of $Fe_3GaTe_2$ and significantly reduce the magnetic transition temperature[30]. By contrast, similar Fe-based bulk vdW ferromagnets $Fe_nGeTe_2$ (n = 3, 4, 5) show enhanced $T_C$ from about 220 K to 310 K with increasing Fe content[31], indicating the important role of Fe content in the ferromagnetism regulation of Fe-based vdW ferromagnets. However, so far, no vdW crystal has exhibited tunable above-room-temperature intrinsic ferromagnetism in the 2D ultrathin regime.

In this work, the tunable intrinsic ferromagnetism in vdW $Fe_{3+x}GaTe_2$ (x = 0 and 0.3) crystals is investigated through magnetic/electrical tests and first-principles calculations. For bulk $Fe_{3+x}GaTe_2$ crystals, the $T_C$ and room-temperature saturation magnetization ($M_S$) can be enhanced up to 376 K and 50.4 emu·g$^{-1}$ by increasing the

Fe content with x = 0.3. For ultrathin $Fe_{3+x}GaTe_2$ nanosheets, the anomalous Hall effect (AHE) and Stoner–Wohlfarth model are used to analyze the $T_C$ and magnetic anisotropy, showing above-room-temperature long-range ferromagnetic order and large room-temperature PMA energy ($K_u$) up to 340 K and $6.6 \times 10^5$ J·m$^{-3}$ in 3-nm $Fe_{3.3}GaTe_2$, respectively, which are the best performance reported in known ultrathin vdW ferromagnetic crystals with the same thickness. First-principles calculations indicate that the additional spin exchange path caused by an increase in Fe content can induce a large magnetic exchange interaction in ultrathin $Fe_{3+x}GaTe_2$ crystals, hence giving rise to the robust intrinsic ferromagnetism and higher $T_C$. This work provides a route to above-room-temperature ferromagnetism in ultrathin vdW magnets.

High-quality $Fe_{3+x}GaTe_2$ crystals with two different Fe contents were grown by chemical vapor transport (CVT) and self-flux methods, respectively (**Note S1**, see details in the **Supplementary Materials**). As shown in **Fig. 1(a)**, $Fe_3GaTe_2$ is a hexagonal vdW material with *P6$_3$/mmc* space group. Electron probe micro-analyzer (EPMA) confirms the elemental composition of as-grown $Fe_{3+x}GaTe_2$ crystals, with Fe molar ratios of 3.03 (denoted as $Fe_3GaTe_2$) and 3.34 (denoted as $Fe_{3.3}GaTe_2$) (**Fig. S1**, see details in the **Supplementary Materials**). Moreover, **Fig. 1(b)** displays the sharp X-ray diffraction (XRD) patterns of $Fe_3GaTe_2$ and $Fe_{3.3}GaTe_2$ single crystals indexed by (00*l*) peaks (*l* is even number), which is consistent with the previous report[3] and indicates high crystallinity. The enlarged (002) peaks present a slight right-shift with increasing the Fe content [right panel of **Fig. 1(b)**], implying the decrease of vdW gap.

According to the Bragg's Law ($2d\sin\theta = n\lambda$, wherein $d$ is the interlayer distance and $n$ is the diffraction order), interlayer distances of $Fe_3GaTe_2$ and $Fe_{3.3}GaTe_2$ crystals are identified as 8.06 Å and 7.83 Å, respectively.

Magnetic property of bulk $Fe_{3+x}GaTe_2$ crystals is measured by vibrating sample magnetometer (VSM). Due to the strong PMA of $Fe_3GaTe_2$ crystals[3], **Figs. 1(c) and 1(d)** exhibit the zero-field-cooling (ZFC) and field-cooling (FC) curves under an out-of-plane magnetic field (B//c-axis) of 500 Oe. Clear ferromagnetism-paramagnetism transitions in first derivation of ZFC curves identify that the $T_C$ of $Fe_3GaTe_2$ and $Fe_{3.3}GaTe_2$ crystal are 354 and 376 K, respectively, indicating a $T_C$ enhancement of about 6.2% by increasing the Fe content [Inset in **Figs. 1(c) and 1(d)**]. The isothermal magnetization curves (M-B) of $Fe_3GaTe_2$ and $Fe_{3.3}GaTe_2$ crystals are different, especially for high temperature regime [**Figs. 1(e) and 1(f)**]. Compared with $Fe_3GaTe_2$, $Fe_{3.3}GaTe_2$ shows a more significant hysteresis loop at 350 K, further confirming a higher $T_C$ of $Fe_{3.3}GaTe_2$ crystals and echoing the results of the ZFC-FC curves. Moreover, the increase in Fe content also enhances the $M_S$ at temperatures from 2 to 350 K (**Fig. S2(a)**, see details in the **Supplementary Materials**). For example, the $M_S$ at 2 K can be tuned from 65.55 to 68.02 emu·$g^{-1}$ (about 3.8% enhancement), which respectively corresponds to a volume magnetization of about 480 and 500 emu·$cm^{-3}$, lager than most $M_S$ = 200 to 340 emu·$cm^{-3}$ of other vdW ferromagnets[32-36]. Meanwhile, the room-temperature $M_S$ shows more significant enhancement from 43.9 to 50.4 emu·$g^{-1}$ (about 14.8% enhancement). Overall, the bulk $Fe_{3.3}GaTe_2$ crystals present record-high $T_C$ and large $M_S$ compared with previous-reported best bulk $Fe_3GaTe_2$ crystals ($T_C$ = 367 K, $M_S$ = 40.11 emu·g$^{-1}$)[3] and other vdW intrinsic ferromagnetic crystals (**Fig. S2(b)**, see details in the **Supplementary Materials**), showing great potential in vdW-integrated spintronics.

To explore the ferromagnetic property of vdW $Fe_{3+x}GaTe_2$ crystals in 2D regime, Hall effect measurement is performed on exfoliated ultrathin $Fe_{3+x}GaTe_2$ (x = 0 and 0.3) nanosheets [**Fig. 2(a)**]. According to the atomic force microscopy (AFM), the thickness of as-tested $Fe_3GaTe_2$ and $Fe_{3.3}GaTe_2$ nanosheets are confirmed as 4 and 3 nm, respectively (**Fig. S3**, see details in the **Supplementary Materials**). Temperature-dependent normalized longitudinal resistance ($R_{xx}$) curves of ultrathin $Fe_3GaTe_2$ and $Fe_{3.3}GaTe_2$ exhibit typical metallic behavior in most temperature ranges, with the latter shows larger $R_{xx}$ at low temperature [**Fig. 2(b)**]. Meanwhile, it should be noted that the upturn of low-temperature $R_{xx}$ in ultrathin $Fe_{3+x}GaTe_2$ nanosheets can be well depicted by the 2D Mott variable-range-hopping (VRH) model[37]:

$$R(T) = R_0 \exp[(\frac{T_0}{T})^{1/3}] \qquad (1)$$

where $R_0$ and $T_0$ are fitting parameters. The fitting curves are presented in **Fig. S4** (see details in the **Supplementary Materials**). Thus, this upturn behavior may be attributed to the emergence of localized electronic states with decreasing temperature, which likely originate from the Fe-3*d* orbitals[3]. Such Fe-3*d* orbitals are also responsible for the itinerant electrons in $Fe_3GaTe_2$ with weak itinerant ferromagnetism[12,38]. Similar behaviors have also been reported in other 2D vdW ferromagnets[10,39] and some disordered 2D transition metal dichalcogenides[40].

For comparing the 2D ferromagnetism in ultrathin $Fe_{3+x}GaTe_2$ with different Fe contents, the AHE is carefully analyzed. Generally, the Hall resistance ($R_{xy}$) of ferromagnetic materials can be divided into ordinary Hall resistance ($R_{OH} = R_0B$) and anomalous Hall resistance ($R_{AH} = R_SM$):

$$R_{xy} = R_0B + R_SM \qquad (2)$$

where B is the component of perpendicular magnetic field, M is the perpendicular magnetization, $R_0$ and $R_S$ are the ordinary Hall and anomalous Hall coefficients, respectively. Due to the metallic behavior and large PMA of $Fe_{3+x}GaTe_2$, the $R_{OH}$ exhibits an extremely small value, which is negligible compared with the $R_{AH}$. Therefore, $R_{xy}$ can be used to represent magnetization. The $R_{xy}$–B curves of ultrathin $Fe_3GaTe_2$ and $Fe_{3.3}GaTe_2$ nanosheets all show clear rectangular hysteresis loops with nearly vertical magnetization flipping till 300 K [**Figs. 2(c)** and **2(d)**]. This indicates the domination of the AHE with a single magnetic domain over the entire 2D crystal, which is reasonable for ultrathin 2D ferromagnets[41]. Together with the large remanent $R_{xy}$ at zero field, these are hallmarks of ferromagnetism with strong PMA[9]. In addition, as the temperature increases, the AHE gradually disappears, which indicates that the 2D $Fe_3GaTe_2$ and $Fe_{3.3}GaTe_2$ crystals go through a transition from a ferromagnetic to a paramagnetic state. Since the AHE is a typical magneto-transport characteristic of materials with intrinsic long-range ferromagnetic orders, the $T_C$ of 4-nm $Fe_3GaTe_2$ and 3-nm $Fe_{3.3}GaTe_2$ can be identified as ~330 and ~340 K, respectively [**Figs. 2(c)** and

**2(d)**]. For few-layer 2D vdW magnets, the $T_C$ will gradually decrease as the thickness decreases[1,2,9-11] due to the disturbance of thermal fluctuations on 2D magnetism. Previous works have demonstrated the layer-dependent $T_C$ of $Fe_3GaTe_2$[11,12], which is consistent with the results of 4-nm $Fe_3GaTe_2$ in this work. Therefore, 3-nm $Fe_{3.3}GaTe_2$ with thinner thickness exhibits a higher $T_C$ than 4-nm $Fe_3GaTe_2$, confirming that the increase in Fe content could effectively enhance the $T_C$ of $Fe_{3+x}GaTe_2$. Importantly, such enhanced high $T_C$ up to 340 K is higher than all other vdW intrinsic ferromagnetic crystals with thickness down to few nanometers[1,2,9-11,31,42-45] [**Fig. 2(e)**], making it a powerful candidate for the practical application of vdW-integrated spintronics with desirable downscaling capability.

In addition, the $H_C$ of 4-nm $Fe_3GaTe_2$ and 3-nm $Fe_{3.3}GaTe_2$ nanosheets at each temperature are extracted in **Fig. S5** (see details in the **Supplementary Materials**), showing the decrease of $H_C$ with increasing temperature. At temperatures below 100 K, the $H_C$ of 3-nm $Fe_{3.3}GaTe_2$ (e.g. $H_C$ = 12.5 kOe at 4 K) is larger than that of 4-nm $Fe_3GaTe_2$ (e.g. $H_C$ = 8.3 kOe at 4 K). When the temperature exceeds 300 K, the $H_C$ of both of them drops below 50 Oe, which is below our sampling interval. This result indicates a temperature-driven transition from hard to soft ferromagnetism for two ultrathin $Fe_{3+x}GaTe_2$ nanosheets.

Another important ferromagnetic property of $Fe_3GaTe_2$ is large room-temperature PMA[3], and thus the influence of Fe content on PMA in ultrathin $Fe_{3+x}GaTe_2$ is carefully discussed. As shown in **Figs. 3(a)** and **3(b)**, $R_{xy}$–B curves of 4-nm $Fe_3GaTe_2$ and 3-nm $Fe_{3.3}GaTe_2$ nanosheets under different field angles ($\theta_B$, the angle between magnetic field and the sample plane) are performed at room temperature. At $\theta_B \approx 90°$, the $R_{xy}$–B curves are square shape due to the large room-temperature PMA. As the field angle decreases, the component of out-of-plane magnetic field gradually decreases (correspondingly, the component of in-plane magnetic field gradually increases), causing the high-field $R_{xy}$ gradually change from an upward trend to a downward trend. A large component of in-plane magnetic field (B ≥ 3 T, $\theta_B \approx 0°$) forces the magnetization to turn towards the in-plane direction, thereby leading to the $R_{xy}$ approaching zero because it is only proportional to the out-of-plane component of the magnetization. The angle between magnetization and the sample plane is defined as the $\theta_M$. The relationship between $\theta_B$ and $\theta_M$ follows the following formula[14]:

$$\theta_M(\theta_B) = \arcsin[\frac{R_{xy}(\theta_B)}{R_{xy}(\theta_B = 90°)}] \qquad (3)$$

**Figs. 3(c)** and **3(d)** present the experimental data of $\theta_M$ as a function of $\theta_B$. Notably, $\theta_M$ is always larger than $\theta_B$ indicates that the magnetization tends to follow the out-of-plane direction regardless of the direction of applying magnetic field, which further echoes the large PMA in two ultrathin $Fe_{3+x}GaTe_2$ nanosheets. Therefore, the magnetization can only be pulled towards the in-plane direction at a sufficiently high magnetic field (e.g. B = 3 T). To evaluate the $K_u$, the Stoner–Wohlfarth model[46] is used to fitting the data and the total energy follows the following formula:

$$E = K_u \sin^2(\theta_M) - \mu_0 H M_S \cos(\theta_B - \theta_M) \qquad (4)$$

Taking the first derivative:

$$\frac{\partial E}{\partial \theta_M} = 2K_u sin(\theta_M)cos(\theta_M) + \mu_0 H M_S \sin(\theta_B - \theta_M) = 0 \tag{5}$$

The experimental data $\theta_M$ and $\theta_B$ are fitted by formula (5), as shown in **Figs. 3(c) and 3(d)**. Assuming the $M_S$ of 4-nm $Fe_3GaTe_2$ and 3-nm $Fe_{3.3}GaTe_2$ are 43.9 and 50.4 emu·g$^{-1}$, the $K_u$ can be obtained as $6 \times 10^5$ and $6.6 \times 10^5$ J·m$^{-3}$, respectively, lager than the $K_u = 3.88 \times 10^5$ J·m$^{-3}$ of previously-reported 9.5-nm $Fe_3GaTe_2$[3]. Enhancing PMA by dimensionality reduction in $Fe_{3+x}GaTe_2$ nanosheets is preferable for stabilizing 2D magnetism[1]. Such large room-temperature $K_u$ is larger than all other vdW ferromagnets and most non-vdW ferromagnetic thin films with thickness of few nanometers[13,47-50] [**Fig. 3(e)**]. Additionally, the similar $K_u$ values of ultrathin $Fe_3GaTe_2$ and $Fe_{3.3}GaTe_2$ indicate that an increase in Fe content does not significantly affect PMA. Therefore, increasing the Fe content in 2D $Fe_{3+x}GaTe_2$ is an effective method for enhancing ferromagnetism, as it can increase $T_C$ with almost no change in magnetic anisotropy, which is desired by ultra-compact spintronic memories and logic devices (e.g. magnetic tunnel junction and spin-orbit torque device) based on ultrathin vdW magnets.

In order to understand the influence of increasing Fe content on intrinsic ferromagnetism and density of states (DOS) of ultrathin $Fe_{3+x}GaTe_2$ crystals, first-principles calculations are conducted on bilayer $Fe_{3+x}GaTe_2$ (x = 0 and 0.25) (**Note S4**, see details in the **Supplementary Materials**). $Fe_3GaTe_2$ and two cases of $Fe_{3.25}GaTe_2$ (denoted as $Fe_{3.25}GaTe_2$-1 and $Fe_{3.25}GaTe_2$-2) are considered in theoretical calculations [**Fig. 4(a)**]. The theoretical interlayer distance of $Fe_3GaTe_2$ (d = 7.85 Å) is larger than that of two $Fe_{3.25}GaTe_2$ cases (d = 7.62 Å), suggesting that the interlayer distance will decrease with increasing the Fe content, which is consistent with the experimental XRD result in **Fig. 1(b)**. Moreover, since the $T_C$ of $Fe_3GaTe_2$ is highly related to the magnetic exchange coupling between neighboring Fe atoms[23,24], the first- and second-nearest exchange parameters ($J_1$, $J_2$), and the additional Fe (denoted as Fe3) induced exchange parameter ($J_3$) have been calculated [**Fig. 4(b)**]. Among them, $J_1$ and $J_2$ are almost insensitive to the increase in Fe content, while the $J_3$ (~22.3 meV) is significant and even very close to the value of $J_1$ (~25.1 meV). Such a large $J_3$ well explains the observed $T_C$ enhancement in ultrathin $Fe_{3+x}GaTe_2$ crystals with increased Fe content.

In addition, bilayer $Fe_3GaTe_2$ and two cases of bilayer $Fe_{3.25}GaTe_2$ all show finite and asymmetric total DOS around the Fermi level, which supports the observed metallic nature and intrinsic ferromagnetism of ultrathin $Fe_{3+x}GaTe_2$ [**Figs. 4(c-e)**] Meanwhile, the *d* orbitals of the Fe play an important role around the Fermi level. By contrast, the contributions of Ga and Te are almost negligible, predominantly from the *p* orbitals of these atoms. Due to the existence of two and three inequivalent Fe atoms in $Fe_3GaTe_2$ and two $Fe_{3.25}GaTe_2$ cases, respectively, further analysis is conducted on the DOS of Fe [**Figs. 4(f-h)**]. It can be found that the total DOS of Fe1, Fe2, and Fe3 mainly comes from their 3*d* orbital electrons, and these Fe atoms all exhibit asymmetric DOS around the Fermi level, implying a certain spin polarization. Moreover, at the Fermi level, the contributions of these three inequivalent Fe atoms to the total DOS are Fe1>Fe2>Fe3.

In conclusion, the above-room-temperature intrinsic ferromagnetism of bulk and ultrathin vdW $Fe_{3+x}GaTe_2$ (x = 0 and 0.3) is investigated. Magnetization tests demonstrate the $T_C$ and $M_S$ enhancement of bulk $Fe_{3+x}GaTe_2$ crystals by increasing the Fe content. For ultrathin $Fe_{3+x}GaTe_2$ nanosheets, the 3-nm $Fe_{3.3}GaTe_2$ shows high $T_C$ of 340 K and large room-temperature $K_u$ of $6.6 \times 10^5$ J·m$^{-3}$ revealed by AHE and Stoner–Wohlfarth model. To the best of our knowledge, these are the best performance in vdW ferromagnetic crystals of the same thickness. Furthermore, first-principles calculations manifest that the asymmetric DOS and an additional magnetic exchange path caused by an increase in Fe content leads to robust intrinsic ferromagnetism and higher $T_C$ in ultrathin $Fe_{3+x}GaTe_2$. This work lays the foundation for above-room-temperature ferromagnetism modulation and ultra-compact spintronic applications based on atomically-thin vdW magnets.

See the supplementary material for the details on the crystal growth, device fabrication, characterizations, first-principles calculations, composition analysis of vdW $Fe_{3+x}GaTe_2$ crystals with different Fe contents by EPMA, comparison of $M_S$ with different Fe contents, temperatures, vdW ferromagnetic bulk crystals, optical images and AFM profile height curves of two devices based on ultrathin $Fe_{3+x}GaTe_2$ nanosheets, $R_{xx}$ measured in two ultrathin $Fe_{3+x}GaTe_2$ nanosheets plotted on a logarithmic (log) scale as functions of $T^{-1/3}$, temperature-dependent $H_C$ from AHE curves of 4-nm $Fe_3GaTe_2$ and 3-nm $Fe_{3.3}GaTe_2$ nanosheets.

This work is supported by the National Natural Science Foundation of China (No. 52272152, 61674063, and 62074061), the National Key Research and Development Program of China (No. 2022YFE0134600), the Natural Science Foundation of Hubei Province, China (No. 2022CFA031), the Shenzhen Science and Technology Innovation Committee (No. JCYJ20210324142010030 and JCYJ20230807143614031), and the Interdisciplinary Research Program of Huazhong University of Science and Technology (No. 2023JCYJ007). The computation is completed in the HPC Platform of Huazhong University of Science and Technology.

## AUTHOR DECLARATIONS

**Conflict of Interest**

The authors have no conflicts to disclose.

**Author Contributions**

Gaojie Zhang and Jie Yu contributed equally to this work.

Gaojie Zhang: Investigation (lead); Methodology (lead); Writing-review and editing (lead). Jie Yu: Investigation (supporting). Hao Wu: Methodology (supporting). Li Yang: Methodology (supporting). Wen Jin: Investigation (supporting). Bichen Xiao: Investigation (supporting). Wenfeng Zhang: Project administration (supporting); Supervision (supporting). Haixin Chang: Conceptualization (lead); Funding acquisition (lead); Project administration (lead); Validation (lead); Writing-review and editing (supporting).

# DATA AVAILABILITY

The data that support the findings of this study are available from the corresponding author upon reasonable request.

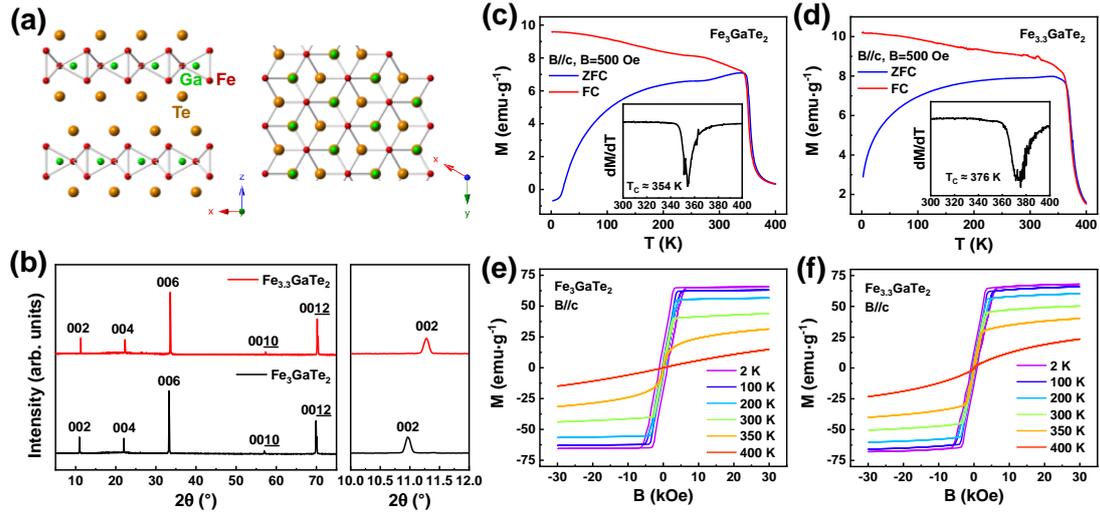

**FIG. 1. Structural and magnetic characterizations of vdW Fe$_{3+x}$GaTe$_2$ (x = 0 and 0.3) crystals.** (a) Schematic diagrams of the crystal structure for vdW Fe$_3$GaTe$_2$. (b) XRD patterns of Fe$_{3+x}$GaTe$_2$ bulk crystals with different Fe contents. Right panel is an enlarge image of (002) peaks. (c, d) ZFC-FC curves of Fe$_3$GaTe$_2$ and Fe$_{3.3}$GaTe$_2$ bulk crystals under out-of-plane magnetic field of 500 Oe. Insets are corresponding first-derivative curves of ZFC curves. (e, f) Isothermal hysteresis loops of Fe$_3$GaTe$_2$ and Fe$_{3.3}$GaTe$_2$ bulk crystals under out-of-plane magnetic field and different temperatures.

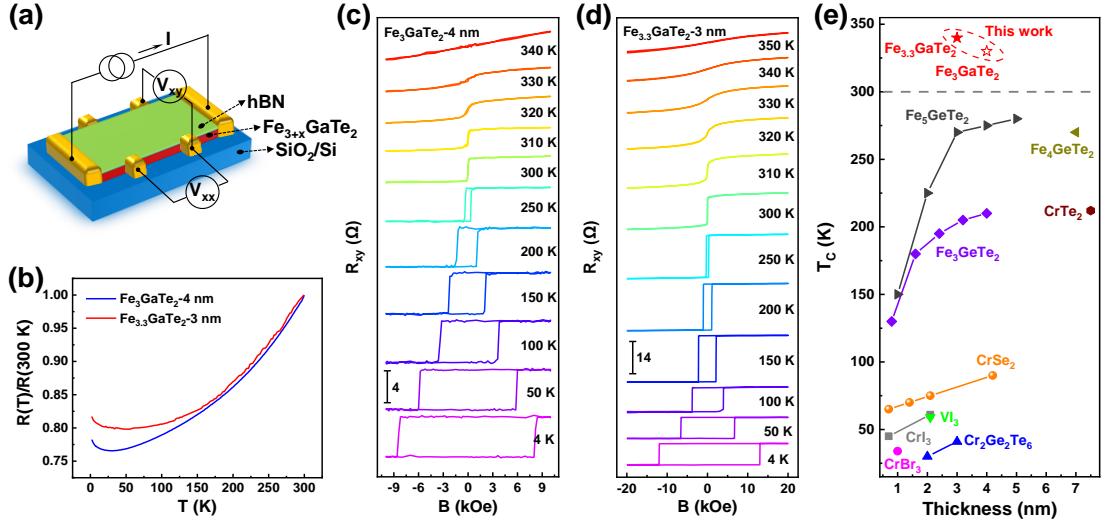

**FIG. 2. Magneto-transport measurements of ultrathin $Fe_{3+x}GaTe_2$ nanosheets.** (a) Schematic diagram of the Hall device measurement setup. (b) Temperature-dependent normalized longitudinal resistance of 2D $Fe_3GaTe_2$ and $Fe_{3.3}GaTe_2$ nanosheets. (c, d) Magnetic field-dependent Hall resistance of 2D $Fe_3GaTe_2$ and $Fe_{3.3}GaTe_2$ nanosheets under different temperatures. (e) Comparison of $T_C$ between ultrathin $Fe_{3+x}GaTe_2$ and other ultrathin vdW intrinsic ferromagnetic crystals with thickness down to few nanometers[1,2,9-11,31,42-45]. Higher $T_C$ and thinner thickness are desirable for ultra-compact 2D spintronics.

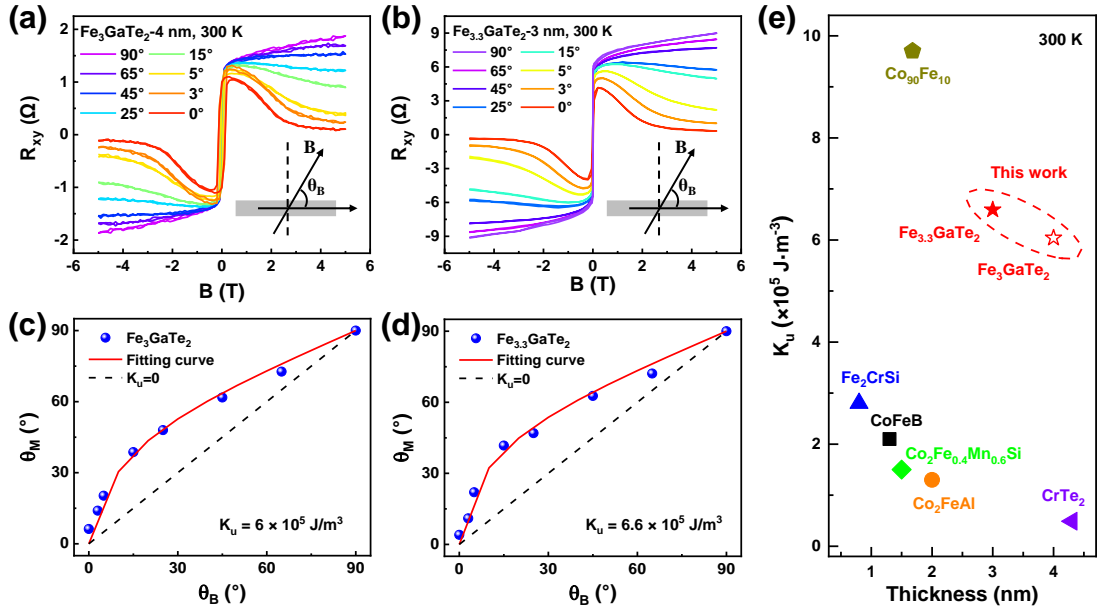

**FIG. 3. Room-temperature magnetic anisotropy analysis of ultrathin $Fe_{3+x}GaTe_2$ nanosheets.** (a, b) Magnetic field-dependent Hall resistance curves of 2D $Fe_3GaTe_2$ and $Fe_{3.3}GaTe_2$ nanosheets under different field angles ($\theta_B$) at 300 K. Insets show the definition of $\theta_B$. (c, d) $\theta_M$ as a function of $\theta_B$. The red solid line is a fit to the Stoner–Wohlfarth model, and the black dash line is $\theta_M = \theta_B$ that corresponds to $K_u = 0$. (e) Comparison of room-temperature $K_u$ between ultrathin $Fe_{3+x}GaTe_2$ and other typical ultrathin ferromagnets with thickness down to few nanometers[13,47-51].

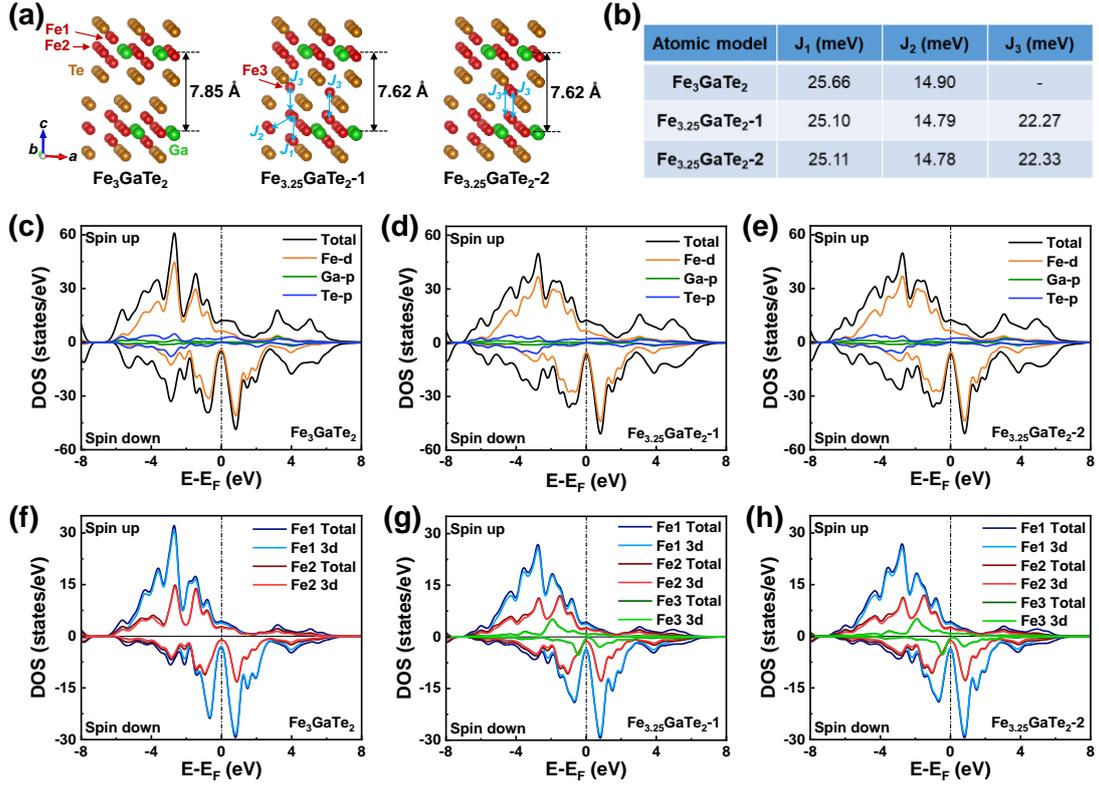

**FIG. 4. First-principles calculations of ultrathin $Fe_{3+x}GaTe_2$ (x = 0 and 0.25).** (a) Atomic models of bilayer $Fe_3GaTe_2$ and two cases of bilayer $Fe_{3.25}GaTe_2$. (b) Calculated magnetic exchange parameters $J_1$, $J_2$ and $J_3$ in each case of bilayer $Fe_{3+x}GaTe_2$. The exchange coupling paths of $J_1$, $J_2$, and $J_3$ are marked by blue arrows in (a). (c-e) Spin-resolved DOS curves of total, Fe $d$, Ga $p$, and Te $p$ in each case of bilayer $Fe_{3+x}GaTe_2$. (f-h) Spin-resolved DOS curves of total Fe1, total Fe2, total Fe3, Fe1 $3d$, Fe2 $3d$, and Te3 $3d$ in each case of bilayer $Fe_{3+x}GaTe_2$. The vertical dashed lines represent the Fermi level.

# Supplementary Information

# Above-room-temperature intrinsic ferromagnetism in ultrathin van der Waals crystal Fe$_{3+x}$GaTe$_2$

Gaojie Zhang[1,2,†], Jie Yu[1,2,†], Hao Wu[1,2], Li Yang[1,2], Wen Jin[1,2], Bichen Xiao[1,2], Wenfeng Zhang[1,2,3], Haixin Chang[1,2,3,*]

[1]State Key Laboratory of Material Processing and Die & Mold Technology, School of Materials Science and Engineering, Huazhong University of Science and Technology, Wuhan 430074, China.

[2]Wuhan National High Magnetic Field Center and Institute for Quantum Science and Engineering, Huazhong University of Science and Technology, Wuhan 430074, China.

[3]Shenzhen R&D Center of Huazhong University of Science and Technology, Shenzhen 518000, China.

[†] Theses authors contributed equally to this work.

[*]Corresponding authors. E-mail: hxchang@hust.edu.cn

**This file includes:**

Experimental Notes 1-4

Supplementary Figures S1-S5

Supplementary References

**Experimental Notes**

**Note S1. Crystal growth**

The $Fe_3GaTe_2$ crystals were grown by chemical vapor transport (CVT) method. High-purity Fe powders (99.95%), GaTe powders (99.99%), and Te powders (99.99%) with molar ratio of 3:1:1 were mixed. Subsequently, the mixture and an amount of the transport agent $I_2$ granules (99.99%) were placed into a long quartz ampoule, then evacuated and sealed. The sealed ampoule was then placed in a two-zone tubular furnace. The temperature of source zone and growth zone were set at 750°C and 700°C for 2 weeks, and then cooled to room temperature naturally.

The $Fe_{3.3}GaTe_2$ crystals were grown by self-flux method. High-purity Fe powders (99.95%), Ga granules (99.999%), and Te powders (99.99%) with molar ratio of 1.1:1:2 were placed into a short quartz ampoule, then evacuated and sealed. The sealed ampoule was then placed in a muffle furnace. The temperature was firstly set at 900°C for 2 days. Subsequently, the temperature was slowly decreased to 780°C within 120 hours, and then cooled to room temperature naturally.

**Note S2. Device fabrication**

A six-terminal Hall electrode with Cr/Au (5/5) was prepared on a $SiO_2$/Si substrate by laser direct writing system, electron beam evaporation, and lift-off process. Then, exfoliated $Fe_{3+x}GaTe_2$ and hBN nanosheets were sequentially dry-transferred to the Hall electrodes. The exfoliation and transfer processes were done in an Ar-filled glove box ($H_2O$, $O_2 \leq 0.1$ ppm) to avoid oxidation.

**Note S3. Characterizations**

The phase and structure were analyzed by powder X-ray diffractometer (XRD, XRD-7000, SHIMADZU) with the Cu K$\alpha$ radiation ($\lambda$ = 0.154 nm). The composition was tested by electron probe micro-analyzer (EPMA, 8050G, Shimadzu). The thickness was tested by atomic force microscope (AFM, Dimension EDGE, Bruker). The magnetic property and magneto-transport measurements were tested by physical property measurement system (PPMS, DynaCool, Quantum Design), and each data point was tested 200 and 25 times and then averaged, respectively.

**Note S4. First-principles calculations**

The density functional theory (DFT)-based first-principles calculations were performed using the Vienna Ab initio Simulation Package (VASP)[1] with the Projector Augmented Wave (PAW) method[2]. The local density approximation (LDA)[3,4] to the exchange-correlation function was used, which is more suitable for describing the magnetic properties of $Fe_3GaTe_2$[5,6]. The cut-off energy of plane wave was set at 450 eV and the convergence criteria for energy and forces were set to $10^{-6}$ eV and 0.001 eVÅ$^{-1}$. The interactions between the slabs were eliminated by setting a 20 Å vacuum layer along the z-axis. Four sublattices ($2 \times 2 \times 1$ supercell) were employed to construct possible magnetic configurations and the Brillouin zone was sampled by a Gamma-centered $9 \times 9 \times 1$ k-point Monkhorst-Pack mesh, which was expanded to $18 \times 18 \times 1$ for the density of states (DOS) calculation. Also, the LSDA plus Hubbard U (LSDA + U) method[7] was employed, and we set the U with 1.5 eV for the Fe 3d electrons.

**Supplementary Figures**

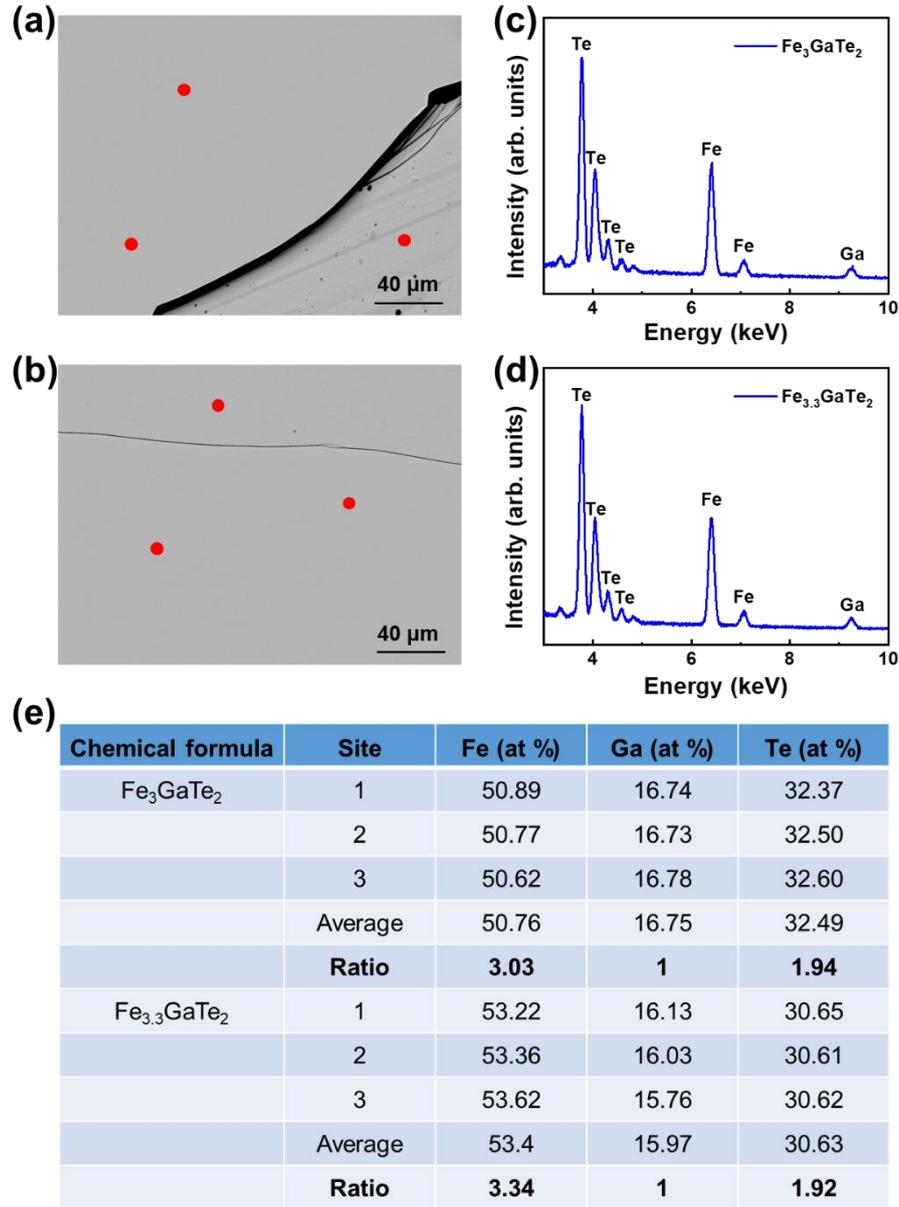

FIG. S1. Composition analysis of vdW Fe$_{3+x}$GaTe$_2$ crystals with different Fe contents by EPMA. (a, b) EPMA images of fresh cleavage Fe$_{3+x}$GaTe$_2$ crystal surface. The red dots indicate the testing micro-zones. (c, d) Typical EDS spectra on the fresh cleavage surface of Fe$_{3+x}$GaTe$_2$ crystals. (e) Elemental percentages collected at three different micro-zones on the fresh cleavage surface of Fe$_{3+x}$GaTe$_2$ crystals.

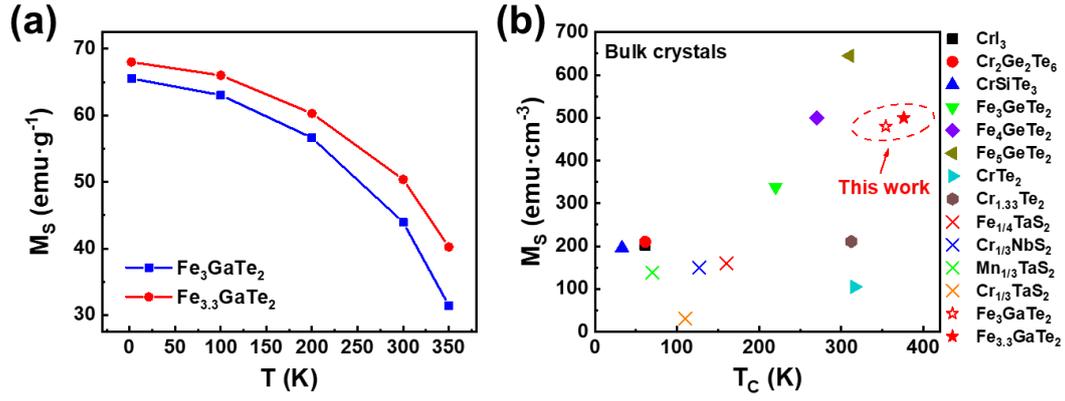

**FIG. S2. Comparison of M$_S$ with different Fe contents, temperatures, and ferromagnetic vdW bulk crystals.** (a) Temperature-dependent M$_S$ extracted from M-B curves of vdW bulk Fe$_3$GaTe$_2$ and Fe$_{3.3}$GaTe$_2$ crystals. The M$_S$ is extracted at 3 T in each M-B curve. (b) Comparison of T$_C$ and M$_S$ of vdW intrinsic ferromagnetic bulk crystals[8-15] and magnetic-atom-intercalated vdW ferromagnetic bulk crystals (cross symbols)[16-19].

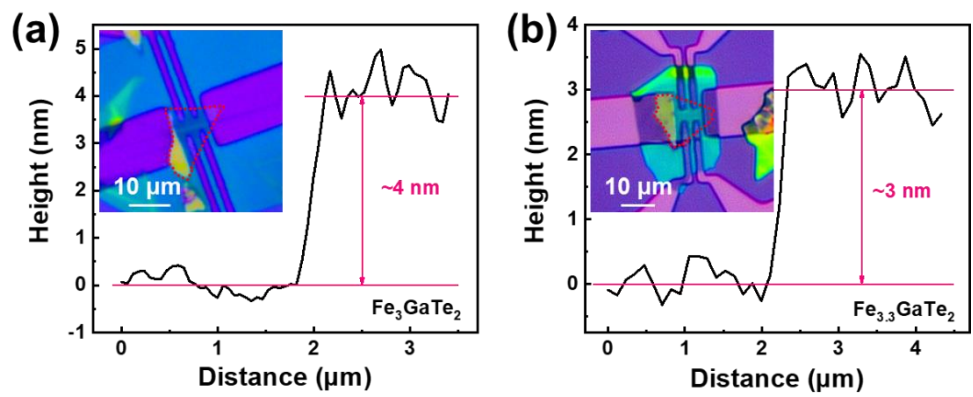

**FIG. S3. Optical images and AFM profile height curves of two devices based on ultrathin $Fe_{3+x}GaTe_2$ nanosheets.** (a) $Fe_3GaTe_2$. (b) $Fe_{3.3}GaTe_2$. The red dash lines in optical images represent the crystal outline.

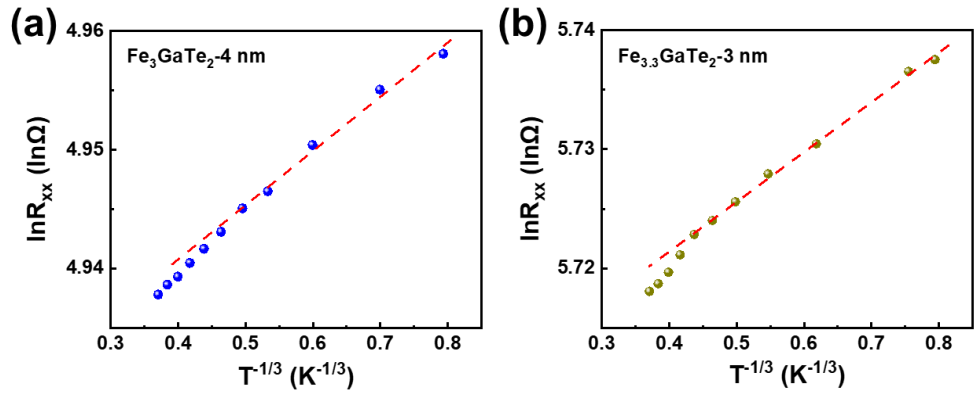

**FIG. S4. $R_{xx}$ measured in two ultrathin $Fe_{3+x}GaTe_2$ nanosheets plotted on a logarithmic (log) scale as functions of $T^{-1/3}$.** (a) $Fe_3GaTe_2$. (b) $Fe_{3.3}GaTe_2$. The red linear fitting curves are based on the 2D Mott VRH model.

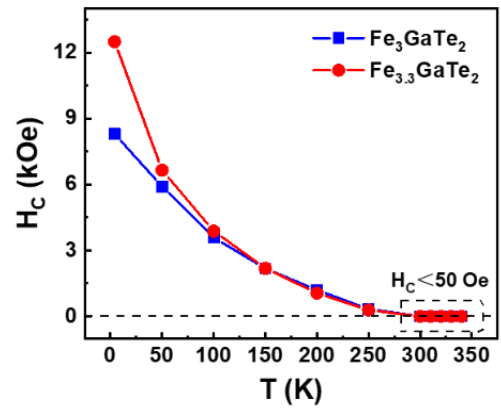

FIG. S5. Temperature-dependent $H_C$ from AHE curves of 4-nm $Fe_3GaTe_2$ and 3-nm $Fe_{3.3}GaTe_2$ nanosheets.

**Supplementary References**